\documentclass[amsmath,amssymb,aps,prl,reprint,floatfix,superscriptaddress]{revtex4-2}

\usepackage{graphicx}
\usepackage{dcolumn}
\usepackage{xcolor}
\usepackage{bm}
\usepackage[version=4]{mhchem}
\usepackage[breaklinks,colorlinks,linkcolor={blue}, %
            citecolor={blue},urlcolor={blue}]{hyperref}

\begin{document}

\title{Midgap state requirements for optically active quantum defects}
\author{Yihuang Xiong} 
\affiliation{Thayer School of Engineering, Dartmouth College, Hanover, New Hampshire 03755, USA}
\author{Milena Mathew}
\affiliation{Department of Electrical Engineering and Computer Sciences, University of California, Berkeley, California 94720, USA}
\affiliation{Materials Sciences Division, Lawrence Berkeley National Laboratory, Berkeley, California 94720, USA}

\author{Sin\'ead M.\ Griffin}
\affiliation{Materials Sciences Division, Lawrence Berkeley National Laboratory, Berkeley, California 94720, USA}
\affiliation{Molecular Foundry Division, Lawrence Berkeley National Laboratory, Berkeley, California 94720, USA}
\author{Alp Sipahigil}
\affiliation{Department of Electrical Engineering and Computer Sciences, University of California, Berkeley, California 94720, USA}
\affiliation{Materials Sciences Division, Lawrence Berkeley National Laboratory, Berkeley, California 94720, USA}
\affiliation{Department of Physics, University of California, Berkeley, California 94720, USA}

\author{Geoffroy Hautier} 
\affiliation{Thayer School of Engineering, Dartmouth College, Hanover, New Hampshire 03755, USA}
\date{\today}

\begin{abstract}
Optically active quantum defects play an important role in quantum sensing, computing and communication. The electronic structure and the single-particle energy levels of these quantum defects in the semiconducting host have been used to understand their opto-electronic properties. Optical excitations that are central for their initialization and readout are linked to transitions between occupied and unoccupied single-particle states. It is commonly assumed that only quantum defects introducing levels well within the band gap and far from the band edges are of interest for quantum technologies as they mimic an isolated atom embedded in the host.  In this perspective, we contradict this common assumption and show that optically active defects with energy levels close to the band edges can display similar properties. We highlight quantum defects that are excited through transitions to or from a band-like level (bound exciton) such as the T center and Se$\rm _{Si}^+$ in silicon. We also present how defects such as the silicon divacancy in diamond can involve transitions between localized levels that are above the conduction band or below the valence band. Loosening the commonly assumed requirement on the electronic structure of quantum defects offers opportunities in quantum defects design and discovery especially in smaller band gap hosts such as silicon. We discuss the challenges in terms of operating temperature for photoluminescence or radiative lifetime in this regime. We also highlight how these alternative type of defects bring their own needs in terms of theoretical developments and fundamental understanding. This perspective clarifies the electronic structure requirement for quantum defects and will facilitate the identification and design of new color centers for quantum applications especially driven by first principles computations.
\end{abstract}

\maketitle

\section{Introduction}
Optically addressable defects in semiconducting hosts can emulate ``artifical atoms" and are central to proposed quantum computing, sensing, and communication  technologies\cite{Wolfowicz2021,Yan2021,Atature2018,Lukin2020}. A series of ``quantum defects" have been identified for quantum information science (QIS) applications including color centers in diamond (e.g., the NV center, the Si divacancy), and SiC (silicon vacancy or the divacancy), as well as the recently emerging defects in silicon such as the T center \cite{Neumann2008,Bergeron2020,Higginbottom2022}. These defects constitute multi-qubit quantum registers based on electron and nuclear spins that can be optically initialized, measured and entangled over long distances\cite{Hensen2015Nature,Bradley2019PRX, Stas2022}.

In tandem with efforts to integrate such defects in devices, there is a growing need to understand, control and predict their fundamental properties from both experimental and theoretical perspectives -- especially with first principles calculations \cite{Dreyer2018,Alkauskas2016,Bassett2019}. These computations can provide information on the defect's electronic structure (more precisely, the single-particle energy levels introduced by the defect with respect to the band edges of the host material) which is essential for predicting many properties of importance for QIS applications such as spin multiplicity, energy of electronic excitations, and radiative lifetimes. 

Understanding a defect's optical excitations is especially important as photons can be used to initialize and read out a defect's electronic spin state. Single-particle levels are commonly used to probe the energy and nature of a defect's optical excitations using transitions within a spin channel from occupied to unoccupied single-particle levels. Early on in the field, an important design criteria on single-particle defect levels emerged for optimal optically-addressable spin-defect qubits \cite{Weber2010,Weber2011,Gordon2013}. This criteria states that both the occupied and unoccupied single-particle levels used during optical excitations need to lie well within the host band gap and far from the band edges. 

This requirement has been used in many recent studies especially using first principles computations to identify new quantum defects.\cite{Varley2016,Tsai2022,Lee2022,Smart2021,Huang2022,Pershin2021, Frey2020, Bathen2021, Ping2021, Cholsuk2022, Smart2021, Narang2019, Tsai2022, Huang2022, Lee2022, Bowes2019, Yeonghun2022} Recent review papers have been often ambivalent on the need of in-gap defect states \cite{Bassett2019,Zhang2020} but some rare reviews have hinted succinctly at how this constraint might not be necessary.\cite{Wolfowicz2021} Importantly, this often-stated need for in-gap defect levels has also been used to rationalize why a wide band gap is a requirement for the host excluding smaller band gap materials such as silicon \cite{Redjem2020}.

The need for localized single-particle defect levels well within the band gap and far from the band edges comes naturally when thinking about mimicking an artificial atom embedded in a solid host \cite{Bassett2019}. It is also a way to reproduce the electronic structure of the NV center in diamond which has been the most widely studied quantum defect. However, this reasonable rationale contradicts the electronic structure of a few well-studied quantum defects. An important requirement for a quantum defect's electronic structure, especially for a spin-photon interface, is the ability to use light to interact with its electronic spin state for initialization and readout. This can indeed be achieved with defects comprising localized states in the gap as per the commonly accepted requirement, but also with alternative electronic structures. In this perspective, we survey a series of defects which are already promising for QIS and highlight that, contrary to the commonly held opinion, electronic structures with a single defect state in the band gap can lead to practical quantum defects. While these single level color centers bring inherent challenges that we will also discuss, this perspective encourages to relax the existing criteria on the electronic structure of quantum defects. This clarification will facilitate the understanding and design of new color centers for QIS especially guided by first-principles computations.

\section{Electronic structure of representative quantum defects in diamond and silicon}

To illustrate the mismatch between the often-stated need for defect levels far from the band edges and the characteristics of \textit{actual} quantum defects, we revisit the electronic structure of four representative defects: the NV center \cite{Neumann2008, Pompili2021} and Si divacancy \cite{Sukachev2017, Sipahigil2014} in diamond, as well as the selenium substitutional defect \cite{DeAbreu2019, Morse2017} and T center in silicon \cite{Bergeron2020, Higginbottom2022, Dhaliah2022,Ivanov2022}. These defects have well-characterized atomic structure and have already shown promise as spin-photon interfaces. We compute their single-particle energies using the hybrid Heyd-Scuseria-Ernzherhoff functional (HSE) which has become the standard for defect calculations \cite{Freysoldt2014, Dreyer2018}. We report both the ground- and excited-state electronic structure, the latter using constrained-HSE where the occupation of unoccupied states is imposed \cite{Jin2021}. Our calculated zero-phonon lines (ZPLs) agree well with experiment for all of the defects considered. We also report the transition dipole moment ($\boldsymbol{\mu}$) for different electronic transitions, which we then use to estimate their radiative lifetime ($\tau$). Full methodological details are provided in the Methods section.

Figure 1 shows the electronic structure of the NV center in its ground and excited state. The single-particle levels obtained in a defective supercell calculation are plotted in red if they are localized on the defect, and in blue if they are delocalized host-like states. In agreement with previous work \cite{Gali2009,Weber2010,Gali2019}, our computations show that the NV center is a triplet state with a possible excitation between the a$_1$(2) and the e$_x$ or e$_y$ states. This excitation of the NV center happens through a strong optical transition ($\mu=4.49$ D) between well-localized levels within the band gap. The computed ZPL is about 2.00 eV, in good agreement with the measured optical transition at 1.945 eV \cite{Davies1976}. It is clear that the electronic structure of the NV center fits exactly the requirement of occupied and unoccupied defects states (a$_1$(2) and e$_x$, e$_y$) within the band gap and far from the band edges.

\begin{figure}[t]
 	\centering
 	\includegraphics[width=0.45\textwidth]{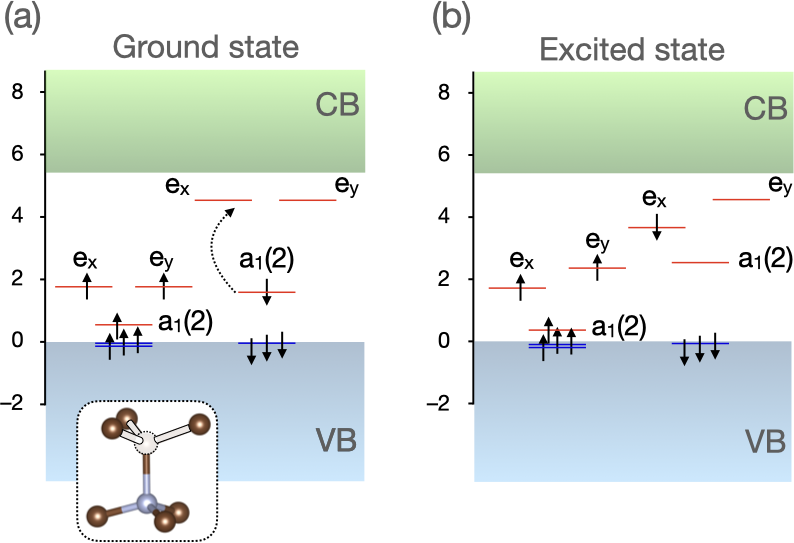}    
 	\caption{Single-particle energy levels of NV center computed using HSE in (a) the ground state and using constrained-HSE in (b) the excited state. The defect structure is shown in the inset of (a).
 	}
 	\label{Fig.1 NV center KS levels}
\end{figure}

A different picture emerges upon examination of more recently identified quantum defects in silicon: substitutional selenium (Se$\rm _{Si}^+$) and the T center ((C-C-H)$\rm_{Si}$) \cite{DeAbreu2019, Morse2017, Bergeron2020, Higginbottom2022, Dhaliah2022}. The Se$\rm _{Si}^+$ is a simple substitutional defect in the silicon matrix and has shown promising properties but emits in the mid-infrared (2902 nm). The T center is a complex defect based on two carbons and a hydrogen substituting a silicon site, and emits in a more technologically attractive wavelength (1326 nm). Both defects have doublet ground states and their electronic structure is shown in Figure 2. The two defects show only one single-particle level in each spin channel. In fact, their lowest energy optical excitation involves a transition between a localized defect state and a host level. For the Se$\rm_{Si}^+$, the transition is between the localized $a_1$ defect state and a conduction band-like state (Figure 2 a,b). 
For the T center, the transition is between a valence band-like state and a localized a'' defect state (Figure 2 c,d).  Both computed ZPLs agree fairly well with experimental data (computed 398 meV vs. measured 427 meV for Se$\rm _{Si}^+$, and computed 985 meV vs. measured 935 meV for T center) \cite{Morse2017, Bergeron2020}. Instead of involving two localized states within the band gap as in the NV center, the excitation of these two silicon defects involves a host-like state. The particle in the host-state (the hole for the T center or the electron for the Se defect) is delocalized and attracted by the defect center forming a so-called defect-bound exciton. The excited state of the T center can be described as a hole bound to a (C-C-H)$\rm_{Si}^{-}$ defect. The excited state of Se$\rm _{Si}^+$ is composed of an electron bound to a Se$\rm_{Si}^{2+}$ defect. Color centers excited through bound excitons have been widely studied in silicon well before the interest in QIS \cite{Davies1989, CLightowlers1985,Steger2011}. In view of the small band gap for silicon, bound exciton color centers might be more common than NV-like defects where the states involved in the excitation are both localized within the band gap. It is striking that the T center and Se$\rm_{Si}^+$ defects -- which are so far the only defects proposed to act as a spin-photon interface in silicon -- are both bound-exciton color centers. In any case, the deviation of these defects' electronic structures from the apparent requirement of localized defect states within the band gap and far from the band edge is clear.

\begin{figure}[t]
 	\centering
 	\includegraphics[width=0.45\textwidth]{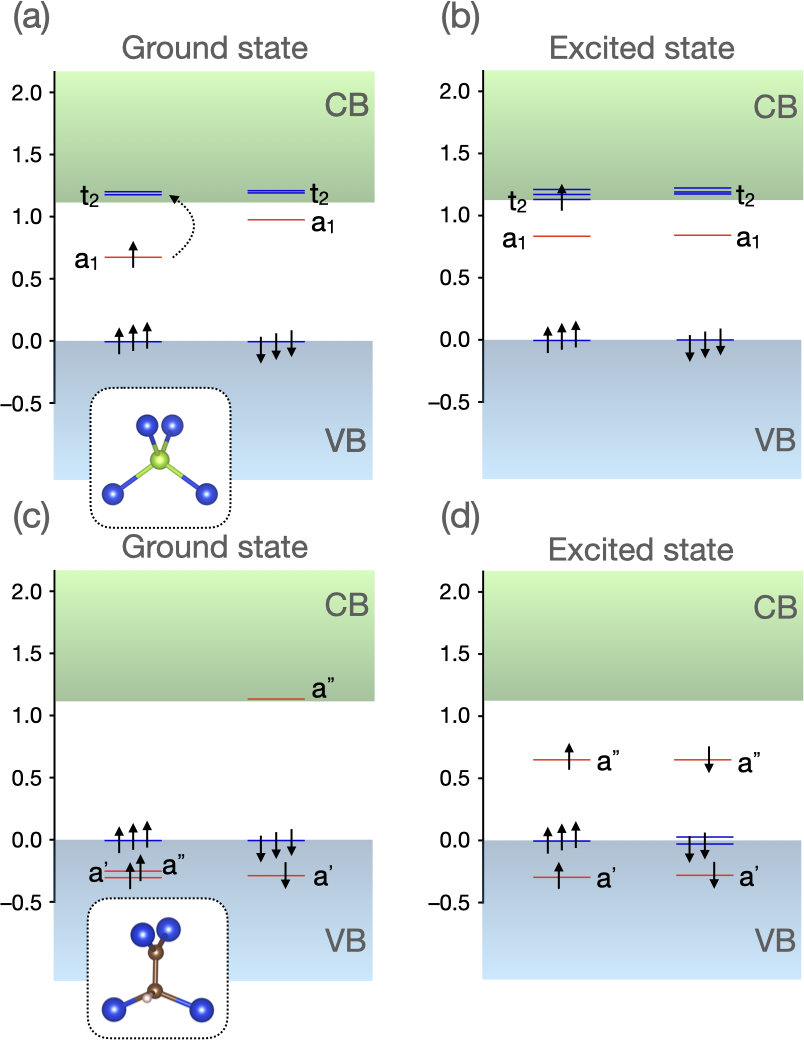}    
 	\caption{Single-particle energy levels of Se$\rm_{Si}^{+}$ (a and b) and the T center (c and d) in silicon using HSE. The excited state is obtained using the constrained-HSE approach. The atomic structure of the defects are shown as the insets.
 	}
 	\label{Fig.2 T center}
\end{figure}

The final defect we consider is the silicon divacancy in diamond. Its centrosymmetric nature precludes the presence of the linear DC Stark shift and makes it less sensitive to spectral diffusion \cite{Sipahigil2014,Udvarhelyi2019}. The silicon divacancy also shows high brightness and long spin coherence time although only at a low temperature (100mK) at low strain \cite{Sukachev2017}. Figure 3 shows the computed electronic structure for the silicon divacancy. Again, we find only one unoccupied single particle state available within the band gap (e$_g$ state). The lowest energy excitation would involve a valence-band like host state suggesting the formation of a bound exciton. However, the transition between the valence band (a$_{1g}$) and e$_g$ is optically forbidden \cite{Gali2013}. The next allowed excitation involves a transition between two localized defect states: from the e$_u$ state sitting 339 meV below the valence band to the unoccupied e$_g$ localized state. This transition leads to a computed ZPL in good agreement with experiment (1.77 eV vs. 1.68 eV) and is much stronger ($\mu=$6.0 D). Our results agree with previous computational results \cite{Gali2013}.

Here again, the canonical design criteria for quantum defects requiring that defect states lie relatively far from the band edges is not satisfied despite the high viability of the silicon divacancy as a quantum defect. In contrast to the defect that induces bound exciton, however, the transition still occurs between localized defect states. The defect excitation is symmetry ``protected" from detrimental radiative recombinations to the band edges states by selection rules and weak optical transitions.

\begin{figure}[t]
 	\centering
 	\includegraphics[width=0.45\textwidth]{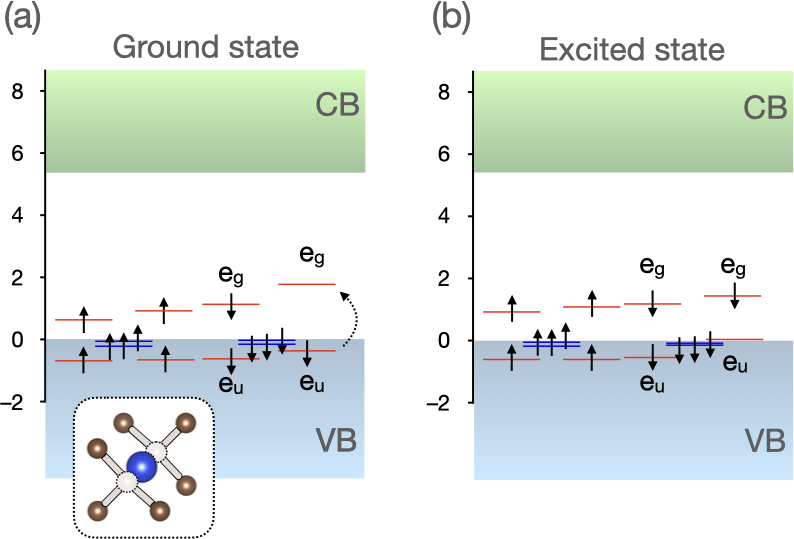}    
 	\caption{Single-particle levels of Si splitting vacancy in diamond using HSE in (a) the ground state and (b) the excited state computed using constrained-HSE. The defect structure is shown in the inset of (a).
 	}
 	\label{Fig.3 Si divacancy}
\end{figure}

\section{Discussion: pros and cons of the quantum defects exhibiting one energy level in the band gap}

From our analysis, we propose the classification of quantum defects into three categories depending on the nature of their optical excitations as described in Figure 4: in-gap intra-defect transitions (e.g., NV center in diamond), bound-exciton transition (e.g., e.g., T center, Se$\rm_{Si}^{+}$), and off-gap intra-defect transition(e.g., Si divacancy in diamond). The last category could also include defects with localized defect states above the conduction band \textit{and} below the valence band. This electronic structure has been recently suggested for the W center in silicon \cite{Baron2022}. 
Experimental evidence from the Si divacancy, T center and Se$\rm_{Si}^{+}$ defects indicates that defects in the last two categories are viable for QIS applications contrary to commonly stated requirements. However, they do have inherent limitations. We evaluate the general trends for each of these cases and resulting trade-offs for QIS-specific properties such as radiative lifetime, brightness and binding energies. We emphasize here that the optimal properties for a quantum defect are application specific.

The proximity of defect levels to the band edges raises the possibility for the excited hole or electron to be captured by bulk host states. For bound-exciton defects such as Se$\rm_{Si}^+$ and the T center, this process can be thought of as the breaking of the bound exciton. Indeed, the delocalized particle (hole or electron) is  relatively weakly bound to the charged defect and can break free through thermal excitation, for instance. For the T center, the breaking of the exciton can be described as a process where the (C-C-H)$\rm_{Si}^{*}$ excited state breaks to form a free hole and a (C-C-H)$\rm_{Si}^{1-}$ charged defect ((C-C-H)$\rm_{Si}^{*}$ → (C-C-H)$\rm_{Si}^{1-}$ + h$^+$). Likewise, for the Se$\rm_{Si}^{1+}$ center, the  exciton can be broken via Se$\rm_{Si}^{1+*}$ → Se$\rm_{Si}^{2+}$ + e$^-$. The binding energy of a defect-bound exciton can be relatively small. For the T center, it is experimentally evaluated to be between 22 and 35 meV \cite{Bergeron2020, Irion1985, Safonov1996}. This implies that the exciton will break and photoluminescence will disappear at a high temperature. In practice this means that the defects that induce bound excitons are limited to operate at cryogenic temperatures only. In contrast, optical control of defects with in-gap intra-defect transitions, such as the NV center, are more feasible for room-temperature operation \cite{Fukami2019, Bao2021}. The silicon divacancy on the other hand, only has a weak temperature dependence of its radiative efficiency \cite{Neu2013, Nguyen2018, Hepp2014}. This suggests that highly localized electronic states inside the valence band and conduction band can be robust against ionization. It is unclear how general that behavior is however. Additionally, we note that the ionization rate of the excited states must be compared against radiative recombination rates to understand the radiative recombination probability. Efficient defect-bound exciton or off-gap transition defects are likely to rely on a favorable radiative rate compared to non-radiative capture by the band edges. The computation of these non-radiative capture process has recently made tremendous progress and are now often performed on electronic and photovoltaic materials \cite{Turiansky2021,Alkauskas2014,Kim2019,Zhang2022}. As new emerging defects moving from the typical in-gap intra-defect transitions are becoming more prevalent in QIS, we expect these theoretical development to play an important role in their understanding and design.

\begin{figure*}[t]
 	\centering
 	\includegraphics[width=0.8\textwidth]{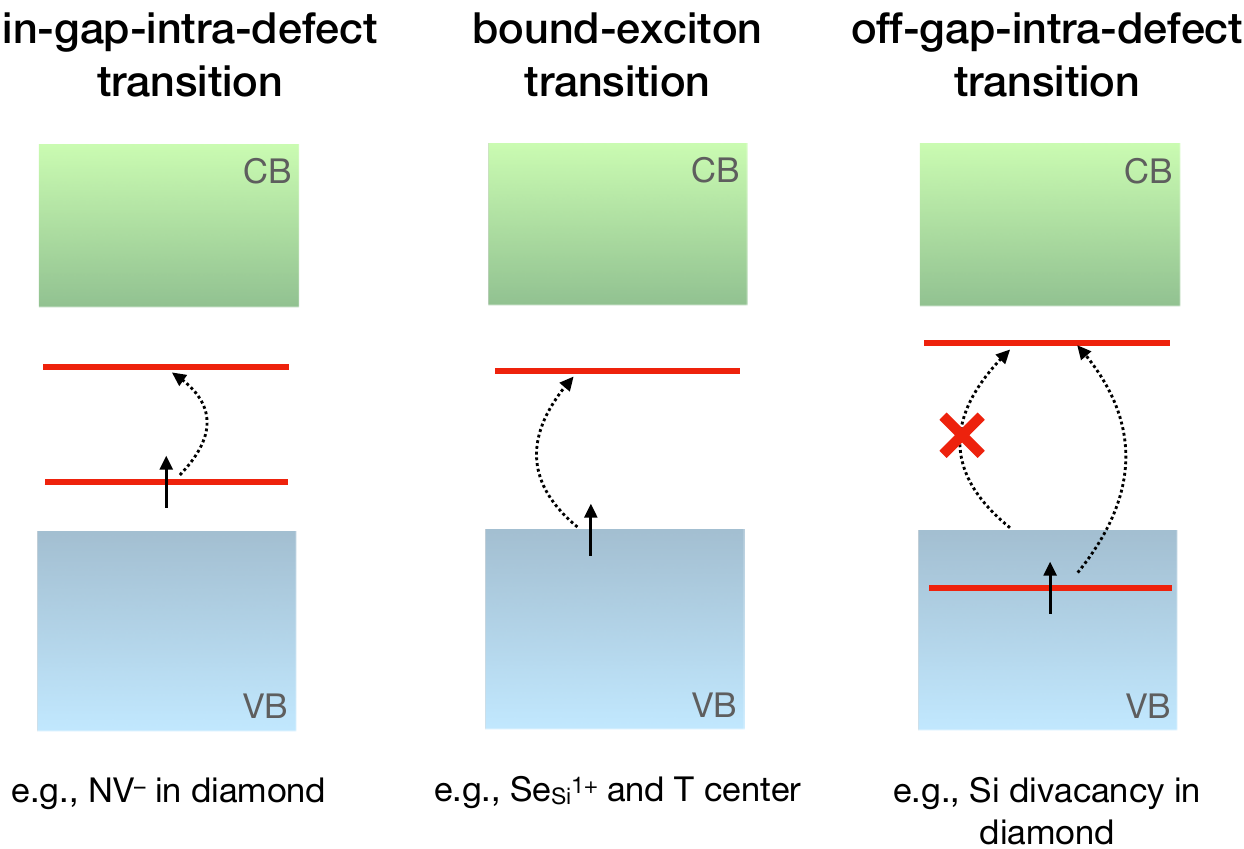}    
 	\caption{Quantum defects can be classified into three categories based on characteristics of their electronic structures. From left to right, there are in-gap intra-defect transitions with both defect levels in the gap, bound-exciton transitions that involve the host state and an in-gap defect state, and off-gap intra-defect transitions. 
 	}
 	\label{Fig.4 Classification of defects}
\end{figure*}

Nonetheless, defects for quantum computing and communication applications require narrow optical transitions which are only achievable at cryogenic temperatures. This is because defects interact with acoustic phonons resulting in level broadening for all color centers studied, including those in diamond \cite{Fu2009,Jahnke2015}. To circumvent such phonon-induced decoherence, quantum communication and computing protocols operate below 10K. However, the need for low temperature operation is a serious drawback for quantum sensing. It is also unclear if a defect excited through a bound exciton process could be designed to exhibit higher exciton binding energy.

Another fundamental issue with the defect-induced bound exciton is their long radiative lifetime. The nature of the optical transition is very different between two localized levels such as those with in-gap intra-defect transitions and defects excited through bound excitons. The transition dipole moment is directly related to the single-particle wavefunctions (initial and final wavefunctions $\psi_i$ and $\psi_f$)  through the transition dipole moment $\boldsymbol{\mu} \propto <\psi_i|\boldsymbol{r}|\psi_f>$. In the case of in-gap intra-defect transitions, the initial and final single-particle states are localized. For the NV center for instance, the wavefunctions are molecular-like with most of their electronic density around the vacancy and the nitrogen. When these transitions are symmetry-allowed, they can have a strong wavefunction overlap, resulting in a large transition dipole moment and a short radiative lifetime (on the order of a few ns). For the bound exciton, one  state is localized and the other is delocalized generally leading to a smaller overlap of the wavefunctions. Both experimental and computed data confirm this picture, with the T center and Se$_{Si}^+$ showing long radiative lifetimes experimentally: 0.94 $\mu$s and 0.9 $\mu$s \cite{Bergeron2020, DeAbreu2019}, respectively, which are in agreement with the respective computed lifetimes of 5.37 $\mu$s and 1.0 $\mu$s. The NV center on the other hand, shows a much shorter lifetime of the order of the ns (both experimentally 12 ns and theoretically 4.4 ns) \cite{Batalov2008, Goldman2015}. Off-gap but intra-defect transitions such as the one in the Si divacancy do not suffer from the lower radiative rate of bound exciton defects as it still involves transition between localized defect states. Experimental measurements of the radiative lifetime of the silicon divacancy are indeed short (around $\sim 10$ ns \cite{Sipahigil2016}) in good agreement with our calculated value of 7.2 ns. 

This longer lifetime for bound excitons is a  more limiting aspect than their low binding energy and low temperature operation requirement. Color centers lacking strong radiative transitions, as demonstrated by the work on Se$\rm _{Si}^+$ and the T center, pose a challenge to achieving coherent optical photons. For quantum  applications, the important figure of merit is the ratio of the transition dipole moment $\boldsymbol{\mu}_t\propto <\psi_i|\boldsymbol{r}|\psi_f>$ and the change in static dipole moment during optical excitation $\boldsymbol{\mu}_s\propto <\psi_i|\boldsymbol{r}|\psi_i>-<\psi_f|\boldsymbol{r}|\psi_f>$ \cite{Udvarhelyi2019}. The first determines the desired radiative transition rates, while the latter determines the amount of optical decoherence due to coupling to electric field noise. While the radiative rate can be enhanced using optical cavities, an intrinsically large ratio of $\mu_t/\mu_s$ is desirable for higher fidelity operation. A second alternative approach is to investigate systems with vanishing $\mu_s$ by symmetry. At this stage, it is unclear if $\mu_s$ will be much different in bound exciton defects and it remains to be seen how much higher transition dipole moments can be pushed when designing new bound exciton defects.

Finally, the limited data available on these few defects does not highlight a fundamental difference in accuracy between different classes of defects and our level of theory (HSE). This might be coincidental though and we expect the defect-bound exciton to bring challenges in modeling due to their involvement of both localized and delocalized states. This might require larger supercells and pose fundamental issues in high accuracy quantum embedding approaches which often focuses on localized defect states in defining an active space \cite{Ma2021, Ma2020, Bockstedte2018}. They could also require a better treatment of the excitonic effects using time-dependent density functional theory (TDDFT) \cite{Jin2021} or solving the Bethe-Salpeter Equation (BSE) \cite{Kirchhoff2022}. Theoretical work has only started on these defect-bound excitons \cite{Zhang_ZH2020} but we hope our perspective will motivate the development of theories that are suitable to their unique nature.

\section{Conclusion}

The search for quantum defects has been strongly influenced by the success of the NV center, and mimicking its electronic structure has led the QIS community to advocate that a quantum defect's electronic structure should present at least two single-particle defect levels within the host band gap and far from the band edges. We have argued in this perspective that this is not a necessary condition, highlighting experimentally promising color centers which do not fulfill this constraint. We believe the search criteria for finding promising quantum defects should be loosened to include bound-exciton and off-gap-intra-defect transitions, but the drawbacks of these single-state defects should be kept in mind. They will more easily suffer from temperature quenching of their photoluminescence and the exciton bound defects will lead in general to longer radiative lifetimes. These constraints bring technological challenges but are not insurmountable. These emerging classes of quantum defects also opens new opportunities in developing our theoretical methods and fundamental understanding. For instance, they motivate the development of excited theories approaches that could treat both the localized and delocalized wavefunctions contributing to the bound exciton or suggest to deepen our understanding of carrier capture processes from defect to band edge states. 

We believe clarifying the requirements for quantum defects is essential as the field moves from understanding existing color centers to predicting novel defects. In the multi-dimensional space of defect properties including excitation energy, ease of process and nanofabrication, spin and optical coherence,  it may be that a defect deviating from the "traditional" in-gap transitions would present a very competitive trade-off in properties. The conclusions of this perspective, while applicable to any host, are especially critical to lower band gap hosts such as silicon. In a small gap host as silicon, defects with well separated levels far from the band edges and emitting in a technologically relevant wavelength are naturally expected to be very rare\cite{Redjem2020}. It is unlikely that these rare defects will also exhibit other favorable properties (coherence, ease of processing). In these cases, looking to alternative types of defects (off-gap transitions and bound-exciton) as it already started with the T center \cite{Bergeron2020,Higginbottom2022} might be essential to find a functioning quantum defect.  

\section{Methods}

The first-principles calculations were performed using Vienna \textit{ab initio} simulation package (VASP) \cite{G.Kresse-CMS96, G.Kresse-PRB96} with the projected augmented wave method (PAW) \cite{Blochl1994}. All calculations were spin-polarized with a plane wave cutoff energy of 520eV. The Heyd-Scuseria-Ernzerhof (HSE) \cite{Heyd2003} functional with 25\% exact exchange was used to describe the electronic structures. All the defects studied in this work were simulated in a 512 atoms supercell and a $\Gamma$-only k-point sampling. The defect structures were optimized at a fixed volume until the forces on the ions were smaller than 0.01 eV/$\AA$. The transition dipole moment was evaluated using the single-particle wavefunction that was calculated at HSE level and further processed with PyVaspwfc code \cite{Zheng2018}. The radiative lifetime was approximated using Wigner-Weisskopf theory of fluorescence: \cite{Gali2019,Alkauskas2016.PRB,Davidsson2020}

\begin{equation}
\frac{1}{\tau}=\frac{n_{r} (2 \pi)^3 \nu^{3}|{ \boldsymbol{\mu}}|^{2}}{3 \varepsilon_{0} h c^{3}},
\end{equation}
where $\tau$ is the radiative lifetime, $n_{r}$ is the refractive index of the host, $\nu$ is the transition frequency corresponding to the energy difference of the Kohn-Sham levels, ${\boldsymbol{\mu}}$ is the transition dipole moment, $\varepsilon_{0}$ is the vacuum permittivity, $h$ is the Planck constant, and $c$ is the speed of light. The transition dipole moment is given as:

\begin{equation}
\boldsymbol{\mu}_{k}=\frac{\mathrm{i} \hbar}{\left(\epsilon_{\mathrm{f}, k}-\epsilon_{\mathrm{i}, k}\right) m}\left\langle\psi_{\mathrm{f}, k}|\mathbf{p}| \psi_{\mathrm{i}, k}\right\rangle,
\end{equation}
where $\epsilon_{\mathrm{i}, k}$ and $\epsilon_{\mathrm{f}, k}$ are the eigenvalues of the initial and final states, $m$ is the electron mass, $\psi_{\mathrm{i}}$ and $\psi_{\mathrm{f}}$ are the initial and final wavefunctions, and $\mathbf{p}$ is the momentum operator.

\section{Acknowledgments}
\begin{acknowledgments}
This work was supported by the U.S. Department of Energy, Office of Science, Basic Energy Sciences 
in Quantum Information Science under Award Number DE-SC0022289. 
This research used resources of the National Energy Research Scientific Computing Center, 
a DOE Office of Science User Facility supported by the Office of Science of the U.S.\ Department of Energy 
under Contract No.\ DE-AC02-05CH11231 using NERSC award BES-ERCAP0020966. 
\end{acknowledgments}

\end{document}